\documentclass[a4paper,11pt]{article}
\usepackage{graphicx}
\usepackage{authblk}
\usepackage{a4wide}
\begin{document}

\title{Perspectives for low energy antiproton physics at FAIR
}


\author[1]{Eberhard Widmann \thanks{ew@antihydrogen.at}
}


\affil[1]{Stefan Meyer Institute for Subatomic Physics, Austrian Academy of Sciences\\
              Boltzmanngasse 3, 1090 Vienna, Austria\\
}


\maketitle

\begin{abstract}
The CRYRING accelerator, previously located at the Manne Siegbahn Laboratory of Stockholm University, has been chosen by the FLAIR collaboration as the central accelerator for the planned facility. It has been modified to allow for high-energy injection and extraction and is capable of providing fast and slow extracted beams of antiprotons and highly charged ions. It is currently being installed at the ESR of GSI Darmstadt where it can be used with highly charged ions. The future possibilities for its use with slow antiprotons will be discussed. 
\begin{description}
\item[keywords]{CRYRING; FLAIR; Highly charged ions; slow antiprotons}
\end{description}

\end{abstract}

\section{Introduction}
\label{intro}
\begin{figure}
\begin{center}
  \includegraphics[width=0.8\textwidth]{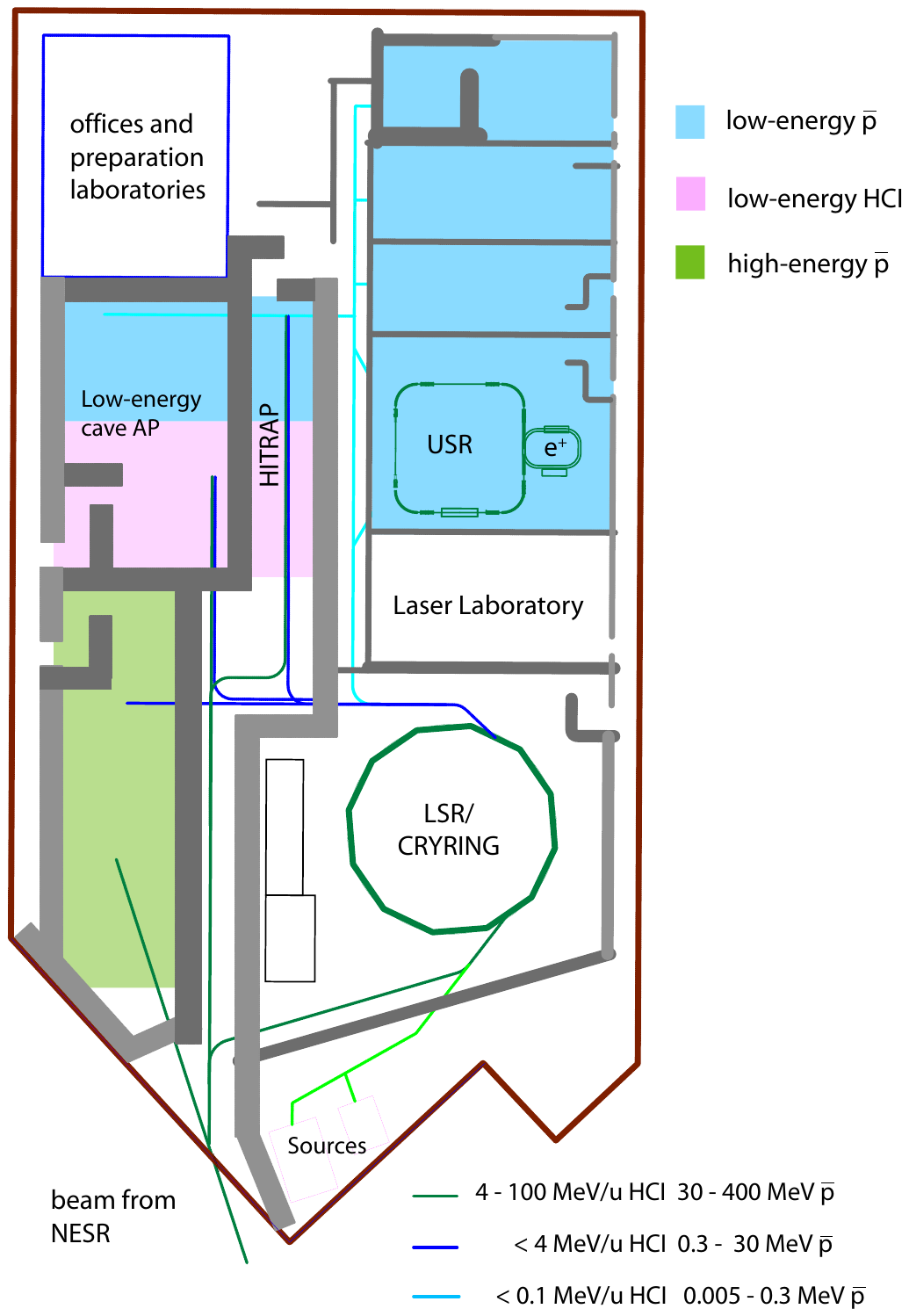}
\end{center}
\caption{The FLAIR facility as proposed in 2005 \cite{FLAIR-TP}.}
\label{fig:FLAIR}       
\end{figure}

\begin{figure*}
  \includegraphics[width=0.8\textwidth]{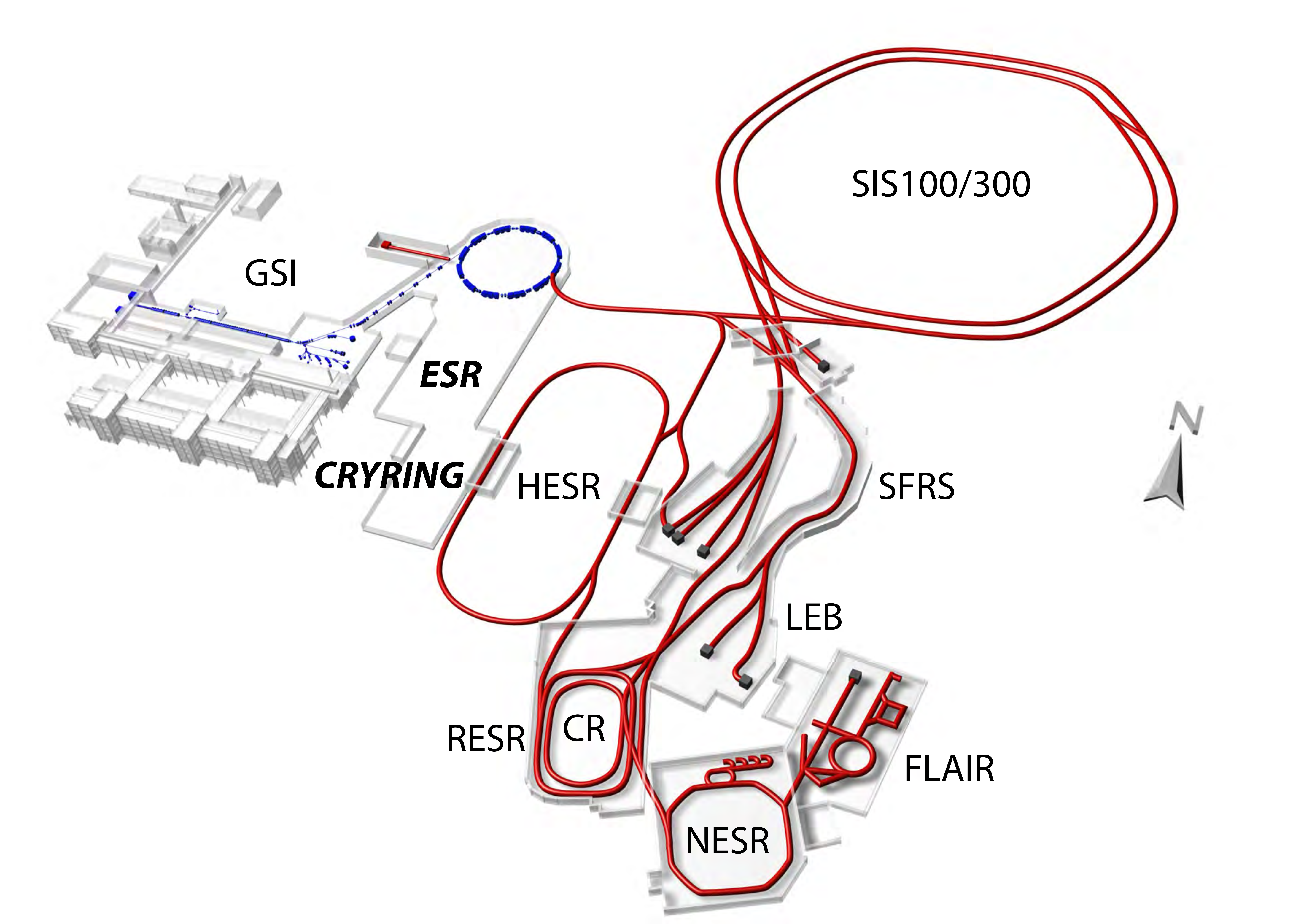}
\caption{Original layout of the FAIR facility showing previously planned NESR and FLAIR buildings to the lower right, and the location of ESR and CRYRING in the existing buildings of GSI.}
\label{fig:FAIR}       
\end{figure*}

In the baseline technical report for the FLAIR facility at FAIR, Darmstadt \cite{FLAIR-TP}, a facility for low-energy antiproton and ion research was described (cf. Fig.~\ref{fig:FLAIR}) that makes use of a magnetic storage ring (Low energy Storage Ring -- LSR) to decelerate antiprotons and highly charges ions and to provide them for further deceleration in an electrostatic storage ring (Ultra-low energy Storage Ring -- USR, minimum energy 20 keV) and a linear decelerator connected to a Penning trap (HITRAP, particles at rest or extracted at a few keV), or to directly use the LSR beam for experiments. The unique features of the LSR are the availability of both slow and fast extracted electron-cooled beams in the energy range 300 keV to 30 MeV for antiprotons. When the Modularized Start Version (MSV) for FAIR was approved, FLAIR together with the NESR (cf. Fig.~\ref{fig:FAIR}) -- which in the TDR concept is needed to decelerate antiprotons and highly charged ions -- was not included.

%
\begin{figure}
\begin{center}
  \includegraphics[width=0.5\textwidth]{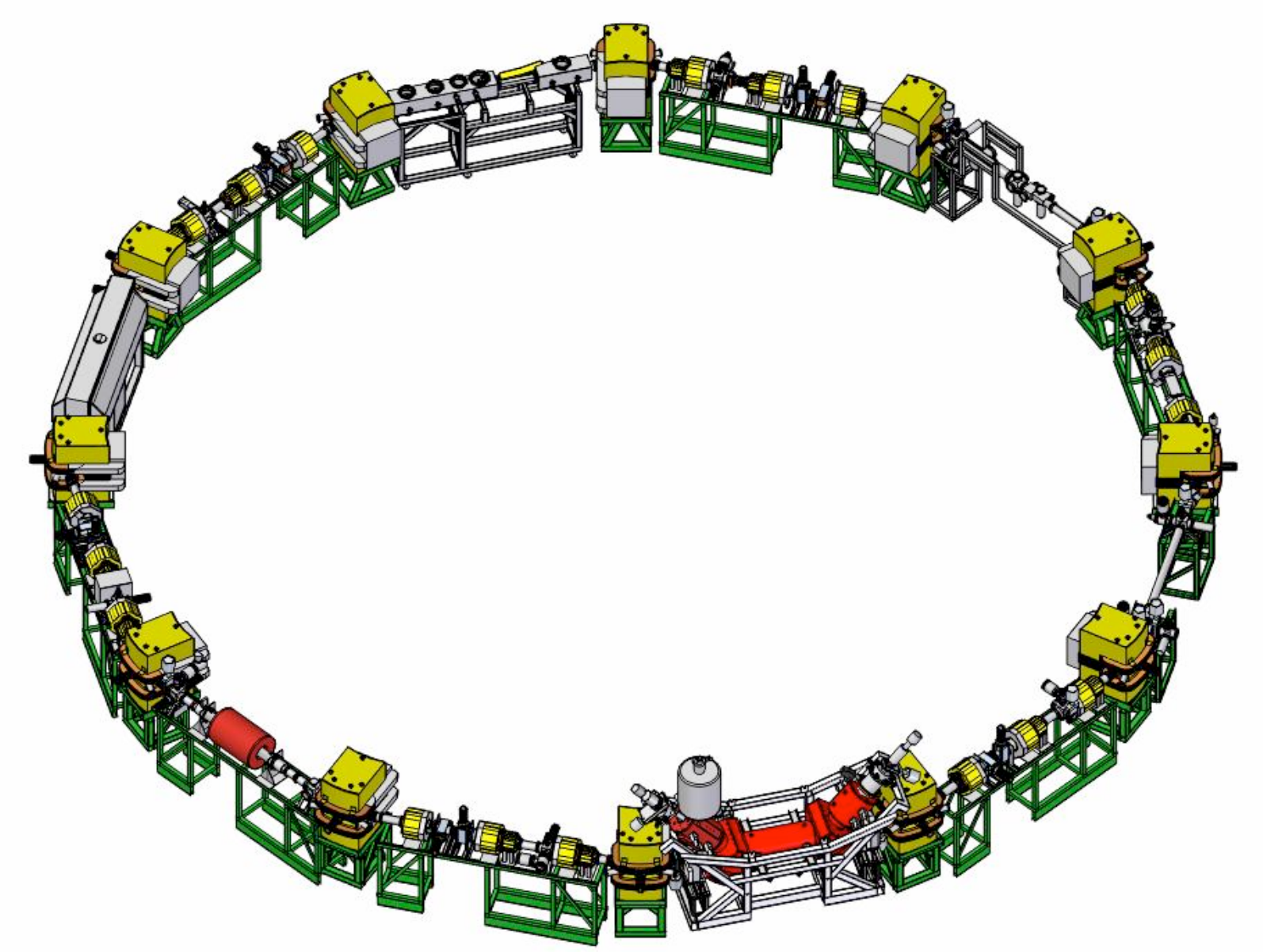}
\end{center}
\caption{Technical drawing of the modified CRYRING \cite{LSR-TDR}.}
\label{fig:1}       
\end{figure}

The specifications for LSR are those for a standard storage ring, and from the very beginning it was obvious that CRYRING, a storage ring that was operated for in-ring atomic physics experiments at the Manne Siegbahn Laboratory (MSL) of Stockholm University \cite{CRYRING}, would be a perfect candidate for LSR. In 2007, after a hearing, the FLAIR collaboration chose to accept CRYRING for the LSR based on a proposal by Sweden to modify the ring to fit the requirement of slow and fast extraction and to provide CRYRING as a Swedish in-kind contribution to FAIR. The MSL team under H{\aa}kan Danared then developed a Technical Design Report \cite{LSR-TDR} which was accepted by FAIR, and subsequently implemented the modifications so that CRYRING became ready for shipment in 2012. 

Based on the readiness of CRYRING and its potential for physics with highly charged ions, an initiative emerged to already install CRYRING at the ESR of GSI rather than keeping it in boxes until the phase 2 of FAIR would be approved. A study group was implemented which produced a report \cite{CRYRINGatESR} showing the scientific potential as well as the technical feasibility. 
 As a result the installation of CRYRING at ESR was approved in autumn 2012 and installation started by now \cite{Herfurt:2013}.


The currently planned location of CRYRING at ESR raises the question on a possibility to provide it with antiprotons from the CR, once the antiproton production target and the CR will be built (in the current planning around 2018). For that several technical questions have to be solved, like the difference in magnetic rigidity $R$ of ESR (maximum $R=10$ Tm) and CR (nominal rigidity $R=13$ Tm). A possible solution here is the production of antiprotons at correspondingly lower energy and their capture in the CR at 10 Tm, which would allow a direct transfer of antiprotons to ESR. Also a beam line to transfer antiprotons from CR to ESR (cf. Fig.~\ref{fig:FAIR}) would be needed. Currently evaluations are in progress about the feasibility for such a transfer line, and the possibility of installing some of the experiments proposed for FLAIR in the ESR experimental hall. As HITRAP is currently also installed at the ESR \cite{Kluge:2008fk}, this could lead to an almost complete realization of the FLAIR facility at this location, going into operation as soon as antiprotons become available at FAIR.



\section{Low-energy antiproton physics experiments with CRYRING}
\label{sec:LEAP}

The physics potential of the FLAIR facility has been outlined in the Letter of Intent \cite{FLAIR-LOI} in 2004 and in \cite{Widmann:2005ys}. It comprises antihydrogen spectroscopy, atomic collision experiments using an internal target in the USR, experiments using antiprotons as hadronic probes, and medical applications. All experiments except the ones making use of USR can be performed also using CRYRING. Since the RESR is not part of the MSV, accumulation is not available, and thus the antiproton rates and lowest energies are similar to the ones available when ELENA at CERN-AD will go into operation in 2017 \cite{ELENA-LEAP2013}. Since CRYRING -- in contrast to ELENA -- will be able to provide slow extraction, experiments of nuclear or particle physics type requiring coincidence techniques would only be possible at CRYRING. If a technical solution and the financing for the transfer line is found, the operation of CRYRING at ESR with antiprotons could start around 2018.

The physics program using cold antihydrogen has been extensively discussed during this conference (see e.g. overviews on ALPHA \cite{ALPHA-LEAP2013}, ATRAP \cite{ATRAP-LEAP2013}, and ASACUSA \cite{HbarHFS-LEAP2013}). In recent years several new experiments at CERN-AD have been approved (AE$\overline{\mathrm{g}}$IS \cite{AEgIS-LEAP2013}, GBAR \cite{GBAR-LEAP2013}, BASE \cite{BASE-LEAP2013}) or are proposed showing a growing interest in low-energy antiproton beams. 

Some of the topics listed int he FLAIR Letter of Intent like the proposal called Exo+pbar \cite{Wada-leap03}, making use of the simultaneous availability of antiprotons and short-lived radio-isotopes, are very promising but would also require the transport of the radio-isotopes created in the CR to CRYRING. 

The availability of  slow extracted antiprotons in an energy range of 300 keV to 30 MeV makes unique nuclear and particle physics experiments possible which were already described in \cite{FLAIR-LOI,Widmann:2005ys}. In addition to the originally proposed production of strangeness $-2$ baryons \cite{Winter:2005vn} in recent years the strangeness production with stopped antiprotons has received renewed interest because of its potential connection to kaonic nuclear bound states that have been predicted \cite{Akaishi:2002qf} and are being searched for in various experiments \cite{KIENLE:2007ys}. Unusually large production rates \cite{Bendiscioli:2007ly} as well as hints for bound states \cite{Bendiscioli:2007dq} were reported from the analysis of LEAR data, making a dedicated search for kaon bound states in $\overline{p}p$ interesting. In addition also bound states with two kaons might be generated \cite{Kienle:2005bh}. Experiments at FLAIR \cite{Zmeskal:2009zr} and for J-PARC using an secondary antiproton beam \cite{Sakuma:2012ve} were proposed and studied. To fully investigate the bound sates a 4$\pi$ detector with the capability of detecting all charged and neutral decay products will be needed.

\section{Summary}

With the installation of CRYRING at ESR a central part of the FLAIR facility becomes available even before FAIR goes into operation. If antiprotons can be brought to the ESR, a major part of the FLAIR facility could be realized around the ESR of GSI. Such a facility is unique because of the availability of slow extracted antiprotons in an energy range of 300 keV to 30 MeV, and could also be used for experiments with trapped antiprotons.

\section*{Acknowledgements}

The author wants to thank \"Orjan Skeppstedt and H{\aa}kan Danared of MSL Stockholm for the close and friendly collaboration that enabled the provision of the unique CRYRING to FLAIR,  Thomas St\"ohlker of GSI for the initiative and support for installing CRYRING at ESR, and the management of GSI and FAIR for their support.

%




%
%

\end{document}